\documentclass{siamonline1116}

\usepackage{lipsum}
\usepackage{amsfonts}
\usepackage{graphicx}
\usepackage{epstopdf}
\usepackage{algorithmic}
\ifpdf
  \DeclareGraphicsExtensions{.eps,.pdf,.png,.jpg}
\else
  \DeclareGraphicsExtensions{.eps}
\fi

\numberwithin{theorem}{section}

\newcommand{\TheTitle}{Trimmed Ensemble Kalman Filter for Nonlinear and Non-Gaussian Data Assimilation Problems} 
\newcommand{\TheAuthors}{W. Li, W. S. Rosenthal, and G. Lin}

\headers{\TheTitle}{\TheAuthors}

\title{{\TheTitle}\thanks{Submitted to the editors DATE.
\funding{This work was supported by the {U.S.} Department of Energy, Office of Science, Office of Advanced Scientific Computing Research, Applied Mathematics program as  part of the Multifaceted Mathematics for Complex Energy Systems (M$^2$ACS) project. G. Lin would like to acknowledge the support from NSF Grants DMS-1555072. Pacific Northwest National Laboratory is operated by Battelle for the DOE under Contract DE-AC05-76RL01830.}}}

\author{
  Weixuan Li\thanks{Advanced Computing, Mathematics \& Data Division, Pacific Northwest National Laboratory, Richland, WA
    (\email{weixuan.li@pnnl.gov}, \email{william.rosenthal@pnnl.gov}).}
  \and
  W. Steven Rosenthal\footnotemark[2]
  \and
  Guang Lin\thanks{Department of Mathematics \& School of Mechanical Engineering, Purdue University, West Lafayette, IN
    (\email{guang.lin@pnnl.gov}).}
}

\usepackage{amsopn}

\ifpdf
\hypersetup{
  pdftitle={\TheTitle},
  pdfauthor={\TheAuthors}
}
\fi

\usepackage[acronym]{glossaries}     
\newacronym{enkf}{EnKF}{ensemble Kalman filter}
\newacronym{tenkf}{TEnKF}{trimmed ensemble Kalman filter}
\newacronym{pf}{PF}{particle filter}
\newacronym{pdf}{PDF}{probability density function}
\newacronym{iid}{i.i.d.}{independent and identically distributed}

\usepackage{subfig}
\usepackage{bbm}

\begin{document}

\maketitle

\begin{abstract}
We study the ensemble Kalman filter (EnKF) algorithm for sequential data assimilation in a general situation, that is, for nonlinear forecast and measurement models with non-additive and non-Gaussian noises. Such applications traditionally force us to choose between inaccurate Gaussian assumptions that permit efficient algorithms (e.g., EnKF), or more accurate direct sampling methods which scale poorly with dimension (e.g., particle filters, or PF). We introduce a trimmed ensemble Kalman filter (TEnKF) which can interpolate between the limiting distributions of the EnKF and PF to facilitate adaptive control over both accuracy and efficiency. This is achieved by introducing a trimming function that removes non-Gaussian outliers that introduce errors in the correlation between the model and observed forecast, which otherwise prevent the EnKF from proposing accurate forecast updates. We show for specific trimming functions that the TEnKF exactly reproduces the limiting distributions of the EnKF and PF. We also develop an adaptive implementation which provides control of the effective sample size and allows the filter to overcome periods of increased model nonlinearity. This algorithm allow us to demonstrate substantial improvements over the traditional EnKF in convergence and robustness for the nonlinear Lorenz-63 and Lorenz-96 models.
%
%
\end{abstract}



\section{Introduction}
\def\real{\mathbb{R}}
\def\hi#1{^{#1}}
\def\lo#1{_{#1}}
\def\mbf#1{{\boldsymbol{#1}}}
\def\of{\circ}
\def\var{\mathrm{var}}
\def\trans{^{\sf T}}
\def\inv{^{-1}}
\def\tsum{\textstyle{\sum}}
\def\proof{\noindent{\sc Proof.} \ }
\def\ali{&\,}
\newcommand{\indep}{\;\, \rule[0em]{.03em}{.67em} \hspace{-.25em}
\rule[0em]{.65em}{.03em} \hspace{-.25em}
\rule[0em]{.03em}{.67em}\;\,}
\def\sd{\mathrm{sd}}
\def\inv{^{-1}}
\def\eop{\hfill $\Box$}
\def\var{\mathrm{var}}
\def\cov{\mathrm{cov}}
\def\cip{\stackrel{P}{\to}}

A sequential data assimilation problem involves estimating the unknown state variables of a dynamic system from a time sequence of measurement data.
From a probabilistic point of view, the solution of such a problem is given by the \textit{posterior} probability density function (PDF) of the states conditioned to the measurement data.
Theoretically, this forms an inverse problem and the required posterior PDF can be derived following the Bayesian filtering approach. In practice, however, a closed-form expression for the posterior PDF does not exist except for a few simple problems. 
Consequently, we have to resort to a numerical algorithm to seek an approximate solution.

The ensemble Kalman filter (EnKF) \cite{evensen1994sequential,evensen2009data} and its variants (e.g., \cite{tippett2003ensemble,li2014adaptive,liao2015data}) are a class of widely used algorithms for sequential data assimilation problems encountered in various scientific and engineering areas, such as atmospheric science \cite{houtekamer2001sequential,ott2004local}, hydrology \cite{reichle2002hydrologic,moradkhani2005dual}, reservoir engineering \cite{gu2005history,aanonsen2009ensemble}, and power systems \cite{li2012application,fan2015dynamic}.
Designed as a Monte Carlo approximation of the classic Kalman filter algorithm, the EnKF employs an ensemble of model simulations to represent the uncertainty associated with the dynamic system under study.
For linear dynamic systems with linear measurement operators and Gaussian noises, the EnKF's solution has been shown to converge, as the ensemble size becomes sufficiently large, to that of the Bayesian filter (the Bayesian filter becomes the classic Kalman filter in this scenario) \cite{mandel2011convergence}.
However, the dynamic models in many practical applications are nonlinear, and the noise follows non-Gaussian distributions.
In this situation, the result given by the EnKF in general does not converge to the correct posterior.
This is in contrast to a fully nonlinear filtering algorithm like the particle filter (PF) \cite{arulampalam2002tutorial,djuric2003particle}, which gives the correct estimation of the posterior PDF, regardless of the linearity and Gaussianity conditions, provided that the number of particles is sufficiently large \cite{crisan2002survey}.

For moderately nonlinear and non-Gaussian problems, the EnKF algorithm can yield satisfactory approximate results with acceptable computational cost, whereas a classic nonlinear data assimilation algorithm such as the PF algorithm may be prohibitively expensive as it requires simulating a much larger number of realizations to overcome the degeneracy issue (that is, the effective number of realizations shrinks rapidly after a few data assimilation steps). Considering the compromise between computational cost and estimation accuracy, the EnKF could be a preferable choice. Nevertheless, when the nonlinearity in the model is strong, and/or the noise is notably non-Gaussian, the EnKF could lead to unacceptable results. For example, it is well recognized (see \cite{tarantola2005inverse} for example) that, to make robust inference from data with outliers, one should adopt a long-tail distribution model rather than a Gaussian distribution model on which the EnKF is built.

There have been some research efforts to combine the advantages of both the EnKF and the PF, for example, by using the Gaussian mixture filter \cite{stordal2011bridging} or via a two-stage hybrid update scheme \cite{frei2013bridging,chustagulprom2016hybrid}. 
In another study, van Leeuwen \cite{van2010nonlinear} applied importance sampling to the particle filter with a proposal transition density based on the Gaussian component of the forecast and observation cross-covariance. Also, Lei and Bickel \cite{lei2011moment} proposed to use the importance sampling method to estimate the statistical moments of the conditional PDF of the state variables (as functions with respect to the measurement variables), which are then used to construct a debiasing scheme for ensemble update.

In this paper, we investigate a method that reduces the bias of the EnKF solution and enhances its robustness in nonlinear/non-Gaussian applications. 
The main novelty in this study is that we prove that the asymptotic limit distribution of the EnKF solution can be written in a special form: the weighted average of the ``shifted" conditional PDFs of the state variables (see Subsection \ref{sec:converge_enkf} for details). 
Based on this observation, we propose to correct the bias of the EnKF by multiplying  the forecast joint PDF (of the state and measurement variables) with a non-negative function, termed the ``trimming function", which essentially adjusts the averaging weights. 
Due to its simplicity and flexibility, the proposed trimmed EnKF (TEnKF) approach is widely applicable to generally nonlinear and non-Gaussian data assimilation problems. 
For instance, the TEnKF does not need the measurement noise to be additive, which is a required condition for the alternative methods mentioned earlier.
We also show the TEnKF methodology to be sufficiently flexible to permit adaptive selection of the trimming function and variable ensemble size in sequential data assimilation steps, which can be exploited to enhance the computational efficiency.

The paper is organized as follows. 
In Section 2, we state the sequential data assimilation problem in a general nonlinear non-Gaussian setting, and derive its solution (i.e., the posterior PDF of the state variables) using the Bayesian filtering approach.
The algorithmic procedures of the EnKF and its asymptotic convergence is reviewed in Section 3. 
In Section 4, we describe the trimmed version of the EnKF algorithm with a discussion of its theoretical and practical aspects. We also provide an implementation with extensions to adaptive trimming and ensemble augmentation.
Section 5 provides several numerical examples which illustrate the limiting distributions, adaptivity, and performance of the TEnKF. Concluding remarks are given in Section 6.

\section{Sequential data assimilation problems}

This section briefly reviews the sequential data assimilation problem and derives its exact solution, known as the Bayesian filter, according to some basic rules of probability. 
The mathematical notation used in this paper follows these conventions: 
an uppercase bold letter (e.g., $\mathbf{Y}$) represents a random vector; 
a lowercase bold letter (e.g., $\mathbf{y}$) represents the values that a random vector takes; 
$p_{\mathbf{Y}}(\mathbf{y})$ represents the PDF of the random vector $\mathbf{Y}$ evaluated at $\mathbf{y}$; 
$p_{\mathbf{XY}}(\mathbf{x}, \mathbf{y})$ represents the joint PDF of the random vectors $\mathbf{X}$ and $\mathbf{Y}$ evaluated at $(\mathbf{x}, \mathbf{y})$;
and $p_{\mathbf{Y}|\mathbf{X}=\mathbf{x}}(\mathbf{y})$ represents the conditional PDF of $\mathbf{Y}$, given the condition that $\mathbf{X}=\mathbf{x}$. Without causing any ambiguity, $p_{\mathbf{Y}|\mathbf{X}=\mathbf{x}}(\mathbf{y})$ is further shortened to $p_{\mathbf{Y}|\mathbf{x}}(\mathbf{y})$. Finally, the scalar special case of each of these terms is represented with regular typeface.

\subsection{Problem statement}
Consider a dynamic system that is described by a forecast model:
\begin{equation}
\label{Eq:forecast}
\mathbf{X}_k = f_k(\mathbf{X}_{k-1}, \mathbf{W}_{k}),\quad k=1,2,...,
\end{equation}
where $\mathbf{X}_{k-1}$ and $\mathbf{X}_k$ are the ($N \times 1$) state vectors at time steps $k-1$ and $k$, respectively. 
The value of the state vector output is uncertain due to both the uncertainty in the previous state and the noisy input vector $\mathbf{W}_k$, which represents the forecast model uncertainty.
Also consider observations $\mathbf{Y}_k$ ($M\times 1$ vector) of the state vector, which may be obtained indirectly through a noisy measurement process, represented by the model
\begin{equation}
\label{Eq:measure}
\mathbf{Y}_k = h_k(\mathbf{X}_k, \mathbf{V}_k),\quad k=1,2,...,
\end{equation}
where $\mathbf{V}_k$ is the measurement noise. 
In this paper, we consider the general situation in which both $f_k(\cdot)$ and $h_k(\cdot)$ could be nonlinear functions, and that $\mathbf{W}_k$ and $\mathbf{V}_k$ could follow non-Gaussian distributions. 

In a sequential data assimilation problem, we need to estimate the state $\mathbf{X}_k$ at each time step $k$ based on all the measurements that have been made up to step $k$: $\mathbf{y}^*_1,..., \mathbf{y}^*_k$, where $\mathbf{y}^*_i$ is the measured value of $\mathbf{Y}_i$. When some measurement data are used to estimate the state, we say these data are ``assimilated" into the model of the dynamic system.


\subsection{Bayesian filter solution}
The solution of the data assimilation problem described above can be formally represented with: $p_{\mathbf{X}_k|\mathbf{y}^*_1,..., \mathbf{y}^*_k}(\mathbf{x}_k)$, i.e., the conditional PDF of $\mathbf{X}_k$ given all measurements available up to and including time step $k$. This density known as the posterior PDF, or the Bayesian filter solution of the data assimilation problem.
This subsection gives a general derivation of this conditional PDF.

\textit{Remark 2.1}: Before moving on to the solution, we note that \eqref{Eq:forecast} and \eqref{Eq:measure} describe a 1\textsuperscript{st}-order Markov chain if the noise vectors $\mathbf{W}_k$ and $\mathbf{V}_k$ at different time steps are independent
, which is a commonly assumed condition in sequential data assimilation problems. 
The Markov property implies that, given the conditional PDF $p_{\mathbf{X}_{k-1}|\mathbf{y}^*_1,..., \mathbf{y}^*_{k-1}}(\mathbf{x}_{k-1})$ obtained in step $k-1$, the estimation of $\mathbf{X}_k$ no longer depends on previous measurements $\mathbf{y}^*_1,...,\mathbf{y}^*_{k-1}$. 
To simplify the notation, we drop the dependence on $\mathbf{y}^*_1,...,\mathbf{y}^*_{k-1}$ in all the conditional PDFs throughout the discussion in the rest of the paper. 
For example, $p_{\mathbf{X}_{k-1}|\mathbf{y}^*_1,..., \mathbf{y}^*_{k-1}}(\mathbf{x}_{k-1})$ is shortened to $p_{\mathbf{X}_{k-1}}(\mathbf{x}_{k-1})$, and $p_{\mathbf{X}_k|\mathbf{y}^*_1,..., \mathbf{y}^*_k}(\mathbf{x}_k)$ is shortened to $p_{\mathbf{X}_k|\mathbf{y}^*_k}(\mathbf{x}_k)$, with the implication that $\mathbf{y}^*_1,...,\mathbf{y}^*_{k-1}$ have been assimilated in previous steps. 
This allows us to restrict our analysis to two neighboring time steps: $k-1$ and $k$.

Following some basic rules in probability theory, the required conditional PDF can be obtained in three steps.

1) Determine the \textit{prior} PDF of $\mathbf{X}_k$ from
\begin{equation}\label{eq:prior_Xk}
    p_{\mathbf{X}_k}(\mathbf{x}_k) = \int p_{\mathbf{X}_{k-1}}(\mathbf{x}_{k-1}) p_{\mathbf{X}_k|\mathbf{x}_{k-1}}(\mathbf{x}_k) d\mathbf{x}_{k-1}.
\end{equation}
This PDF is termed \textit{prior} in the sense that it represents the estimation of $\mathbf{X}_k$ prior to the assimilation of measurement data $\mathbf{y}^*_k$. 
Note that in deriving Eq.~\eqref{eq:prior_Xk}, we have used the \textit{sum rule}, i.e., the \textit{marginal} PDF with respect to a variable $x_k$ can be calculated by summing/integrating the \textit{joint} PDF over the other random variables, or
\[
p_{\mathbf{X}_k}(\mathbf{x}_k) = \int p_{\mathbf{X}_{k-1}\mathbf{X}_k}(\mathbf{x}_{k-1},\mathbf{x}_k) d\mathbf{x}_{k-1},
\]
as well as the \textit{product rule}, i.e., the joint PDF equals the product of the marginal PDF and the conditional PDF:
\[
p_{\mathbf{X}_{k-1}\mathbf{X}_k}(\mathbf{x}_{k-1},\mathbf{x}_k) 
= 
p_{\mathbf{X}_{k-1}}(\mathbf{x}_{k-1})
p_{\mathbf{X}_k|\mathbf{x}_{k-1}}(\mathbf{x}_k).
\]
Eq.~\eqref{eq:prior_Xk} shows that the prior PDF $p_{\mathbf{X}_k}(\mathbf{x}_k)$ can be calculated from two pieces of information: the posterior PDF of $\mathbf{X}_{k-1}$ obtained in the previous time step, and the conditional PDF $p_{\mathbf{X}_k|\mathbf{x}_{k-1}}(\mathbf{x}_k)$, also known as the \textit{transition} PDF, which is defined by the forecast model Eq.~\eqref{Eq:forecast}.

2) Find the \textit{joint} prior PDF of $\mathbf{X}_k$ and $\mathbf{Y}_k$ by another application of the product rule:
\begin{equation}\label{eq:joint_pdf}
    p_{\mathbf{X}_k \mathbf{Y}_k} (\mathbf{x}_k, \mathbf{y}_k) = p_{\mathbf{X}_k}(\mathbf{x}_k) p_{\mathbf{Y}_k|\mathbf{x}_k}(\mathbf{y}_k),
\end{equation}
where $p_{\mathbf{X}_k}(\mathbf{x}_k)$ is obtained in Step 1), and $p_{\mathbf{Y}_k|\mathbf{x}_k}(\mathbf{y}_k)$ is defined by the measurement model Eq.~\eqref{Eq:measure}.

3) Compute the required posterior PDF by fixing $\mathbf{y}_k = \mathbf{y}^*_k $ in the joint PDF \eqref{eq:joint_pdf} and re-normalizing the PDF such that its integral over $\mathbf{x}_k$ equals one:
\begin{equation}\label{eq:bayes_rule}
    p_{\mathbf{X}_k|\mathbf{y}^*_k}(\mathbf{x}_k)
    =
    c 
    p_{\mathbf{X}_k}(\mathbf{x}_k)
    p_{\mathbf{Y}_k|\mathbf{x}_k}(\mathbf{y}^*_k),
\end{equation}
where $c = \left[ \int p_{\mathbf{X}_k}(\mathbf{x}_k)
p_{\mathbf{Y}_k|\mathbf{x}_k}(\mathbf{y}^*_k) d\mathbf{x}_k \right] ^{-1}$ is the normalizing constant, and $p_{\mathbf{Y}_k|\mathbf{x}_k}(\mathbf{y}^*_k)$ is known as the \textit{likelihood}.

\section{Ensemble Kalman filter}
\label{sec:enkf}

The posterior PDF given by the Bayesian filter, i.e., Eqs.~\eqref{eq:prior_Xk}-\eqref{eq:bayes_rule}, in general cannot be explicitly and analytically expressed when the forecast and measurement models are nonlinear, or when the distributions of the noises are non-Gaussian. 
Instead, this PDF may be approximately represented with the empirical distribution of $n$ sample points, i.e., a Monte Carlo solution. 
One commonly used method to generate such a sample is EnKF. 
In this section, we review the EnKF algorithm with a discussion on its convergence. This algorithm solves a sequential data assimilation problem through iterations of the two-step process, first sampling from the prior PDF (known as the forecast step in the literature), and then sampling from the posterior PDF (known as the update step) at each time step $k$. These time steps coincide with times when the system is measured.

\subsection{Sample from the prior in the forecast step of the EnKF}
\label{sec:EnKF_forecast}
In the forecast step of the EnKF at time step $k$, $n$ sample points from the joint prior PDF $p_{\mathbf{X}_k \mathbf{Y}_k} (\mathbf{x}_k, \mathbf{y}_k)$ can be obtained by simulating $n$ realizations of the forecast model Eq.~\eqref{Eq:forecast}:
\begin{equation}\label{eq:prior_sample_x}
\mathbf{X}^i_k = f_k(\mathbf{X}^i_{k-1}, \mathbf{W}^i_{k}), \quad i=1,...,n,
\end{equation}
and the measurement model Eq.~\eqref{Eq:measure}:
\begin{equation}\label{eq:prior_sample_y}
\mathbf{Y}^i_k = h_k(\mathbf{X}^i_k, \mathbf{V}^i_k),\quad i=1,...,n,
\end{equation}
where $\mathbf{X}^i_{k-1}$ are independent and identically distributed (i.i.d.) sample points from $p_{\mathbf{X}_{k-1}}(\mathbf{x}_{k-1})$ (estimated in the previous time step), and $\mathbf{W}^i_k$ and $\mathbf{V}^i_k$ are i.i.d. sample points from their respective distributions (given as part of the problem statement).

\subsection{Sample from the posterior in the update step of the EnKF}
\label{sec:EnKF_update}

\textit{Remark 3.1}: Since every random quantity involved in the update step is associated with time step $k$, we drop the subscript $k$ throughout the rest of the paper to further simplify the notation. 
For example, $\mathbf{X}_k$, $\mathbf{y}^*_k$, and $p_{\mathbf{X}_k|\mathbf{y}^*_k}(\mathbf{x}_k)$ are shortened to $\mathbf{X}$, $\mathbf{y}^*$, and $p_{\mathbf{X}|\mathbf{y}^*}(\mathbf{x})$, respectively, without further clarification.

While sampling from the prior PDF in the forecast step is rather straightforward, sampling from the posterior PDF $p_{\mathbf{X}_k|\mathbf{y}^*_k}(\mathbf{x}_k)$ in the update step is tricky because of the extra condition that needs to be satisfied: $\mathbf{Y} = \mathbf{y}^*$. 
Indeed, the distinct feature of the EnKF algorithm in comparison with other sample-based data assimilation methods (e.g., a PF) is its linear update scheme, which is shown below:
\begin{equation}\label{eq:kalman_update}
\tilde{\mathbf{X}}^i=\mathbf{X}^{i} + \mathcal{K} (\mathbf{y}^* - \mathbf{Y}^i),
\quad
i=1,...,n,
\end{equation}
where $\tilde{\mathbf{X}}^i$ approximately follows the target posterior PDF (we will explain the approximation in the next subsection).
${\mathcal{K}} = {\mathcal{C}}_{\mathbf{XY}} {\mathcal{C}}_{\mathbf{YY}}^{-1}$ is the Kalman gain, where ${\mathcal{C}}_{\mathbf{XY}}$ is the covariance matrix between random vectors $\mathbf{X}$ and $\mathbf{Y}$, and ${\mathcal{C}}_{\mathbf{YY}}$ is the covariance matrix of random vector $\mathbf{Y}$. 

\textit{Remark 3.2}: In many EnKF studies, it is assumed that the measurement error is additive: $\mathbf{Y} = h(\mathbf{X}) + \mathbf{V}$, and the error $\mathbf{V}$ is independent of $\mathbf{X}$. 
Thus, the covariance matrices needed in the calculation of the Kalman gain become ${\mathcal{C}}_{\mathbf{XY}} = {\mathcal{C}}_{\mathbf{X}h}$ and ${\mathcal{C}}_{\mathbf{YY}} = {\mathcal{C}}_{hh} + \mathcal{R}$, where ${\mathcal{C}}_{\mathbf{X}h}$ is the covariance matrix between $\mathbf{X}$ and $h(\mathbf{X})$, ${\mathcal{C}}_{hh}$ is the covariance matrix of $h(\mathbf{X})$, and $\mathcal{R}$ is the covariance matrix of $\mathbf{V}$ (assumed to be known).
Additive measurement noise also implies that the update scheme \eqref{eq:kalman_update} becomes
\begin{equation}\label{eq:kalman_update_additive}
\tilde{\mathbf{X}}^i=\mathbf{X}^{i} + {\mathcal{K}} [\mathbf{y}^* - (h(\mathbf{X}^i) + \mathbf{V}^i)].
\end{equation}
Note that often in EnKF literature the update scheme is given in a different form:
\begin{equation}\label{eq:kalman_update_additive2}
\tilde{\mathbf{X}}^i=\mathbf{X}^{i} + {\mathcal{K}} [(\mathbf{y}^* + \mathbf{V}^i )- h(\mathbf{X}^i) ],
\end{equation}
that is, the error realizations $\mathbf{V}^i$ are added to the measurement value $\mathbf{y}^*_k$, which is the procedure known as measurement perturbation \cite{burgers1998analysis}. 
When the error follows a symmetric PDF, i.e., $p_\mathbf{V} (\mathbf{v}) = p_\mathbf{V} (-\mathbf{v})$, \eqref{eq:kalman_update_additive} and \eqref{eq:kalman_update_additive2} are equivalent statistically. 
In a general situation, we use the update scheme \eqref{eq:kalman_update} since it assumes neither additive nor symmetrically distributed measurement noise.

\textit{Remark 3.3}: In general nonlinear and non-Gaussian problems, the covariance matrices $\mathcal{C}_{\mathbf{XY}}$ and $\mathcal{C}_{\mathbf{YY}}$ cannot be analytically derived, and thus, the Kalman gain $\mathcal{K}$ is not known exactly. 
In practice, we can use the sample estimates $\hat{\mathcal{C}}_{\mathbf{XY}}$ and $\hat{\mathcal{C}}_{\mathbf{YY}}$ calculated from the forecast ensemble. 
Then $\mathcal{K}$ may be approximated  with
\begin{equation}
\label{eq:sample_gain}
\hat{\mathcal{K}}=\hat{\mathcal{C}}_{\mathbf{XY}} \hat{\mathcal{C}}_{\mathbf{YY}}^{-1},
\end{equation}
and $\tilde{\mathbf{X}}^i$ may be approximated with $\hat{\mathbf{X}}^i=\mathbf{X}^{i} + \hat{\mathcal{K}} (\mathbf{y}^* - \mathbf{Y}^i)$.
By the law of large numbers, as the sample size $n \to \infty$, we have that $\hat{\mathcal{K}}$ and $\hat{\mathbf{X}}^i$ converge \textit{in probability} to $\mathcal{K}$ and $\tilde{\mathbf{X}}^i$, respectively.

\subsection{Convergence analysis of EnKF update scheme}
\label{sec:converge_enkf}
It has been well-recognized that the sample points given by the EnKF update scheme \eqref{eq:kalman_update} need not converge to the correct posterior PDF asymptotically for the general nonlinear/non-Gaussian data assimilation problem. 
Intuitively, this is because the Kalman gain is calculated only from covariance matrices instead of using the full information described by the joint prior PDF.
The proposition below gives the limit PDF (as the ensemble size $n \to \infty$) of the EnKF sample. Particularly, we show that this limit PDF is equal to the weighted average of the ``shifted" conditional PDFs conditioned to different observation values. This result provides an insight about how to correct the asymptotic bias in the EnKF algorithm.

\begin{proposition}
\label{proposition:enkf asymptotic general}
The PDF of $\tilde{\mathbf{X}} = \mathbf{X} + \mathcal{K} (\mathbf{y}^* - \mathbf{Y})$ is 
\begin{equation}
\label{eq:limit_enkf}
p_{\tilde{\mathbf{X}}} (\tilde{\mathbf{x}})= 
\int 
p_{\mathbf{X}|\mathbf{y}} (\tilde{\mathbf{x}} - \mathcal{K} (\mathbf{y}^* - \mathbf{y}))
p_{\mathbf{Y}}(\mathbf{y})
d\mathbf{y}.
\end{equation}

\end{proposition}
\noindent{\sc Proof.} See Appendix \ref{app:proof1}.
\hfill $\Box$

Proposition \ref{proposition:enkf asymptotic general} shows that the limit PDF  of the EnKF solution can be obtained by first shifting the conditional PDF $p_{\mathbf{X}|\mathbf{y}} (\mathbf{x})$ by $\mathcal{K} (\mathbf{y}^* - \mathbf{y})$ for different conditioning values of $\mathbf{y}$, and then averaging the shifted conditional PDFs with respect to the weighting distribution $p_{\mathbf{Y}}(\mathbf{y})$, i.e., the marginal PDF of $\mathbf{Y}$.
Note that when $\mathbf{y} = \mathbf{y}^*$, the shifted conditional PDF $p_{\mathbf{X}|\mathbf{y}} (\tilde{\mathbf{x}} - \mathcal{K} (\mathbf{y}^* - \mathbf{y}))$ becomes $p_{\mathbf{X}|\mathbf{y}^*} (\tilde{\mathbf{x}} )$, which is exactly our target posterior PDF, the Bayesian filter solution for the general nonlinear/non-Gaussian problem. 
However, for other values of $\mathbf{y}$, the shifted conditional PDFs in general are different from the target posterior, and hence, their weighted average is not guaranteed to equal the target posterior. 
One exceptional situation is when $\mathbf{X}$ and $\mathbf{Y}$ follow a joint Gaussian distribution (usually this is not true if the forecast or the measurement models are nonlinear, or if the noises are non-Gaussian), for which we can easily prove that the shifted conditional PDF $p_{\mathbf{X}|\mathbf{y}} (\tilde{\mathbf{x}} - \mathcal{K} (\mathbf{y}^* - \mathbf{y})) =
p_{\mathbf{X}|\mathbf{y}^*} (\tilde{\mathbf{x}} )$ for any value of $\mathbf{y}$, and as a result, the limit PDF of the EnKF solution is exactly the target posterior.

\section{A trimmed EnKF for nonlinear and non-Gaussian problems}
In the previous section, we investigated the convergence of the EnKF algorithm and discussed its bias from the exact posterior PDF. 
This section introduces a trimming procedure that reduces the bias. We first give the theoretical analysis of the trimming procedure in Subsection \ref{sec:trimming_theory} and then discuss its practical implementation in the following subsections.

\subsection{Bias reduction with a trimming procedure}
\label{sec:trimming_theory}
Proposition \ref{proposition:enkf asymptotic general} shows that the limit PDF of the EnKF solution is an average of the shifted conditional PDFs, weighted by the marginal PDF of $\mathbf{Y}$.
Motivated by this observation, we propose a modification of the EnKF algorithm that allows us to adjust the averaging weight, and thus reduces the bias in the posterior estimate.

The implementation steps of the new algorithm are parallel to that of EnKF described in Sections \ref{sec:EnKF_forecast} and \ref{sec:EnKF_update} except that we now introduce an adjusted joint PDF obtained by multiplying a non-negative function $t(\mathbf{y})$ to the original joint PDF $p_{\mathbf{XY}}(\mathbf{x},\mathbf{y})$:
\begin{equation}
\label{eq:adjust_joint}
    p_{\mathbf{XY}}^t(\mathbf{x},\mathbf{y}) = c_t 
    t(\mathbf{y})
    p_{\mathbf{XY}}(\mathbf{x},\mathbf{y}),
\end{equation}
where $c_t$ is the normalizing constant that ensures the integral of the adjusted joint PDF equals $1$. 
Suppose we can draw sample points $(\mathbf{X}^i_t, \mathbf{Y}^i_t)$ following the adjusted joint PDF \eqref{eq:adjust_joint}. Then, similar to \eqref{eq:kalman_update}, the updated sample points are obtained by
\begin{equation}\label{eq:t_kalman_update}
\tilde{\mathbf{X}}^i_t = \mathbf{X}^i_t + \mathcal{K} (\mathbf{y}^* - \mathbf{Y}^i_t).
\end{equation}

Similar to \eqref{eq:limit_enkf}, we have the following proposition regarding the PDF of the sample points generated with \eqref{eq:t_kalman_update}.
\begin{proposition}
\label{proposition:tenkf asymptotic general}
The PDF of $\tilde{\mathbf{X}}_t = \mathbf{X}_t + \mathcal{K} (\mathbf{y}^* - \mathbf{Y}_t)$, where $\mathbf{X}_t$ and $\mathbf{Y}_t$ follow the adjusted joint prior \eqref{eq:adjust_joint}, is 
\begin{equation}
\label{eq:limit_tenkf}
p^t_{\tilde{\mathbf{X}}} (\tilde{\mathbf{x}})=
\int 
p_{\mathbf{X}|\mathbf{y}} (\tilde{\mathbf{x}} - \mathcal{K} (\mathbf{y}^* - \mathbf{y}))
p_{\mathbf{Y}}(\mathbf{y}) c_t t(\mathbf{y})
d\mathbf{y}.
\end{equation}

\end{proposition}
\noindent{\sc Proof.} See Appendix \ref{app:proof2}.
\hfill $\Box$

Comparing \eqref{eq:limit_tenkf} with \eqref{eq:limit_enkf}, we see that the averaging weight has become $p_{\mathbf{Y}}(\mathbf{y}) c_t t(\mathbf{y})$, which can be adjusted by choosing different functions $t(\mathbf{y})$. 
Particularly, to reduce the bias between the limit PDF \eqref{eq:limit_tenkf} and the true posterior $p_{\mathbf{X}|\mathbf{y}} (\mathbf{x})$,
we want to put more weight on the values of $\mathbf{y}$ that are close to the measurement $\mathbf{y}^*$, and less weight on the values of $\mathbf{y}$ that are far from this measurement. 
Thus, we choose $t(\mathbf{y})$ such that it is monotonically decreasing with respect to the distance between $\mathbf{y}$ and $\mathbf{y}^*$. 
For example, one may choose 
\begin{equation}
\label{eq:trimming_func}
t(\mathbf{y}) = \exp [- d( \mathbf{y} , \mathbf{y}^* ) / \lambda],
\end{equation}
where $\lambda$ is a positive constant, and $d (\mathbf{y} , \mathbf{y}^* )$ is a measure of the distance between $\mathbf{y}$ and $\mathbf{y}^*$. 
For instance, we choose the L1 distance (normalized with the prior sample standard deviation of each state variable) in the numerical experiments in our study: 
\begin{equation}
d (\mathbf{y} , \mathbf{y}^* ) = \sum_{j=1}^N \frac{| y_j -y_j^*|}{\hat{\sigma}_j}.
\end{equation}
Intuitively, multiplying such functions as $t(\mathbf{y})$ to the original joint prior leads to a partial ``trimming off" of the density distribution in the regions that are inconsistent with the measurement data. Hence, we refer to $t(\mathbf{y})$ as the ``trimming function", and the modified version of the EnKF algorithm as the trimmed EnKF, or TEnKF.

One can control how much to trim by adjusting the trimming function. Consider $t(\mathbf{y}) = \exp [-d(\mathbf{y} , \mathbf{y}^* )/\lambda]$ for example.
A large $\lambda$ results in a mild trim. 
In the extreme case, as $\lambda \to \infty$, then $c_t t(\mathbf{y}) \to 1$ for any $\mathbf{y}$ and thus implies zero trim. 
From \eqref{eq:limit_tenkf} we see that $p^t_{\tilde{\mathbf{X}}} (\mathbf{x}) \to p_{\tilde{\mathbf{X}}} (\mathbf{x})$ in this situation, i.e., the limiting posterior distribution of the TEnKF converges to that of the EnKF. 
On the other hand, a small $\lambda$ results in a significant trim. 
In this extreme case, as $\lambda \to 0$, then $c_t t(\mathbf{y}) \to \delta(\mathbf{y}^*-\mathbf{y})/p_{\mathbf{Y}}(\mathbf{y}^*)$ (i.e., the Dirac-delta function), which exemplifies the maximum possible trim. 
From \eqref{eq:limit_tenkf} we see that $p^t_{\tilde{\mathbf{X}}} (\mathbf{x}) \to p_{\mathbf{X}|\mathbf{y}^*} (\mathbf{x})$, i.e., the true posterior PDF for the Bayesian filter problem.
We will demonstrate how the choice of trimming function interpolates between the limiting distribution of the EnKF and the true posterior for a simple test problem in Subsection \ref{subsec:limit_dist_example}.

\subsection{Sampling from trimmed joint prior}
\label{sec:sample_trim}
We discussed in the previous subsection how the trimming procedure affects the limit distribution of the EnKF. 
In this and the following subsections, we focus on the practical implementation details and give the complete description of the TEnKF algorithm.

The first task in implementing the TEnKF algorithm is to sample from the trimmed joint prior PDF \eqref{eq:adjust_joint}. 
A straightforward way to achieve such a sample is implementing an importance sampling procedure using the untrimmed joint prior as the proposal, that is, we associate each sample point realization $(\mathbf{X}^i, \mathbf{Y}^i)$ obtained with  \eqref{eq:prior_sample_x} and \eqref{eq:prior_sample_y} a weight proportional to the trimming function:
\begin{equation}\label{eq:weight}
w_i = \frac{t(\mathbf{Y}^i)}{\sum_{j=1}^n t(\mathbf{Y}^j)}.
\end{equation}
To generate equally weighted sample points, we further apply a bootstrapped resampling to the weighted sample points using these weights as selection probabilities. In other words, for each $i = 1,..., n$, the probability the $i$\textsuperscript{th} member of the trimmed sample is taken to be the $j$\textsuperscript{th} member of the untrimmed sample is $w_j$, and duplicates in the trimmed sample are permitted.

\subsection{Adaptive selection of trimming function}
\label{sec:adaptive_trim}


A critical step in the implementation of TEnKF is the selection of the trimming function $t(\mathbf{y})$ that results in a satisfactory balance between accuracy and efficiency.
In practice, the selection may be made from a family of functions with a tuning parameter that controls the level of trim.
For instance, we can choose from the family $t(\mathbf{y};\lambda) = \exp [- d(\mathbf{y} , \mathbf{y}^* )/\lambda]$ by tuning the parameter $\lambda$.
As discussed in Subsection \ref{sec:trimming_theory}, a larger trim (i.e., a smaller $\lambda$) helps reduce the bias in the TEnKF estimate. 
However, in practice, a larger trim may also lead to sample degeneracy---similar to that of the PF---as a large portion of the ensemble members are given negligible weights and trimmed off.
To deal with this trade-off, we design an adaptive algorithm to automatically tune the trimming function such that the filter maintains a sufficient effective ensemble size $n_e$ after the trimming.

For $n$ weighted ensemble members, the effective ensemble size can be measured by
\begin{equation} \label{eqn:neff}
n_e =  \left[ \sum_{i=1}^n w_i^2 \right]^{-1} ,
\end{equation}
where  $w_i$, as defined in Eq.~\eqref{eq:weight}, are the trimming weights corresponding to a parameter value of $\lambda$.  Note that $n_e$ equals $n$ for untrimmed (equally weighted) ensemble members, and decreases as $\lambda$ becomes smaller (larger trim). In practice, we perturb $\lambda$ in some iterative scheme until $n_e$ is close to a target effective ensemble size $n_e^*$. 

The advantage of this algorithm is that it allows adaptive tuning of the trimming function based on available computational resources. We summarize this procedure in Algorithm \ref{alg:tr}, which maintains a specified effective ensemble size by automatically adjusting the tuning parameter. When only a relatively small number of ensemble members can be simulated, the algorithm can enforce a mild trim (large $\lambda$), and thus behave more like the EnKF. On the other hand, if we can simulate an ensemble size that is significantly larger than the target $n_e^*$, the algorithm will automatically adopt a significant trim (small $\lambda$) to reduce the bias in the posterior estimate. We will illustrate this adaptive behavior in Subsection \ref{subsec:nonlin}.

\begin{algorithm}
\caption{TEnKF \label{alg:tr}}
\begin{enumerate}
\item Given the (untrimmed) forecast ensemble $(\mathbf{X}^i, \mathbf{Y}^i)$ , $i=1,...,n,$ the measurement value $\mathbf{y}^*$, the distance measure function $d(\cdot,\mathbf{y}^*)$, and the trimming parameter $\lambda$:
\item Calculate the Kalman gain $\hat{\mathcal{K}}$ with \eqref{eq:sample_gain}.
\item \label{alg:tr:weight} Compute the weights $w_i$ for the ensemble members with \eqref{eq:trimming_func} and \eqref{eq:weight}.
\item Compute the effective ensemble size $n_e$ with \eqref{eqn:neff}.
\item Decrease/increase $\lambda$ if $n_e$ is greater/smaller than the target $n_e^*$, until $n_e \approx n_e^*$.
\item Obtain the trimmed ensemble $(\mathbf{X}^j_t, \mathbf{Y}^j_t)$ , $j=1,...,n,$ by bootstrapped resampling of $(\mathbf{X}^i, \mathbf{Y}^i)$ with respect to the trimming weights $w_i$.
\item Compute the updated state ensemble $\tilde{\mathbf{X}}^i_t$, $i=1,...,n$, with \eqref{eq:t_kalman_update}.
\end{enumerate}
\end{algorithm}

\subsection{Adaptive ensemble sizing}
\label{subsec:AdaptEns}

Other metrics can be applied alongside the trimming function to further control the effective ensemble size and to enhance the computational efficiency. We note that the level of nonlinearity/non-Gaussianity in the model often varies over different data assimilation cycles. An efficient algorithm should deploy more resources when needed to maintain accuracy and prevent degeneracy, and operate with less computational cost during more linear/Gaussian intervals. We take advantage of the flexibility of the TEnKF and propose adaptively increasing the ensemble size $n$ prior to the trimming step. Specifically, we examine the number of forecast realizations that are within a distance $d_{\mathrm{max}}$ from the observations:
\begin{equation}
\label{eq:n_d}
n_d = \sum_{i=1}^n \mathbbm{1}_{d(\mathbf{Y}^i , \mathbf{y}^*) < d_{\mathrm{max}}}.
\end{equation}
where the indicator function $\mathbbm{1}_A$ equals $1$ whenever statement $A$ is true, and $0$ when not. Then before trimming (immediately prior to Step \ref{alg:tr:weight} in Algorithm \ref{alg:tr}), we increase the forecast ensemble size up to $n_{\mathrm{aug}} = \lfloor n \min\left( r_{\mathrm{max}}, n/n_d \right) \rfloor$, where $r_{\mathrm{max}}$ is a cap on the augmentation ratio $n_{\mathrm{aug}}/n$. 
This step is intended to increase the effective ensemble size, as measured by Eq.~\eqref{eq:n_d}, to approximately the original ensemble size $n$.
We examine the efficacy of this strategy used in consort with trimming in Subsection \ref{subsec:nonlin}.

\section{Numerical examples}

We demonstrate the properties and efficacy of the TEnKF and the aforementioned algorithms in several numerical examples. Using two well-known numerical models due to Lorenz \cite{lorenz63,lorenz96}, we illustrate how Proposition \ref{proposition:tenkf asymptotic general} implies the TEnKF posterior estimate interpolates between those of the EnKF and PF in the large $n$ limit. Additional numerical exercises show that the TEnKF restores the convergence of the EnKF with increasing ensemble size $n$ as model nonlinearities increase. We also examine how well adaptive control of ensemble size prior to trimming allows the filter to push through transient nonlinearities without sacrificing accuracy.

\subsection{Limiting distributions}
\label{subsec:limit_dist_example}
\begin{figure}[!t] \centering
\vspace{0pt}
\includegraphics[height=2in,clip=true,trim=0 0 20 15]{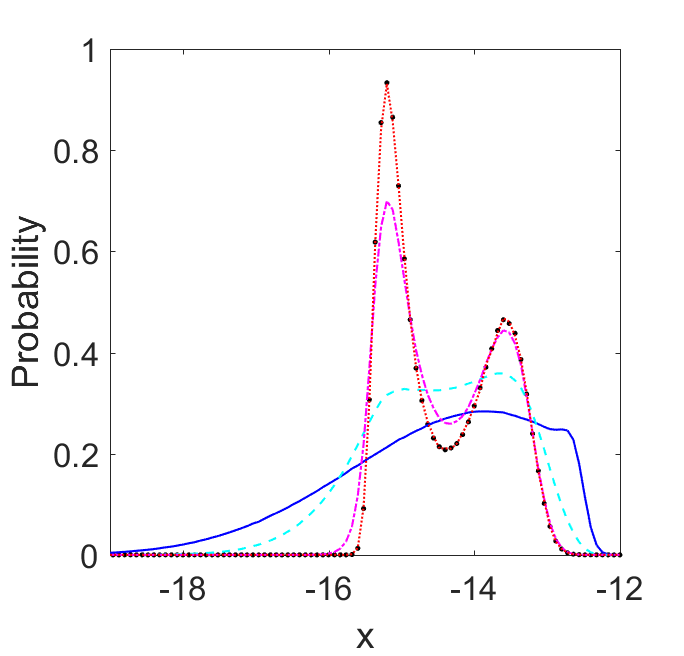}
\vspace{0pt}
\includegraphics[height=2in,clip=true,trim=10 0 15 15]{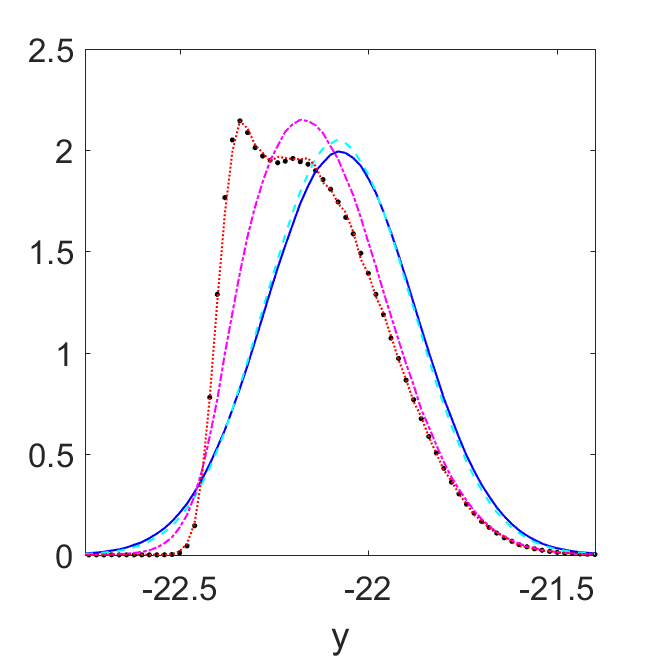}
\vspace{0pt}
\includegraphics[height=2in,clip=true,trim=0 0 20 15]{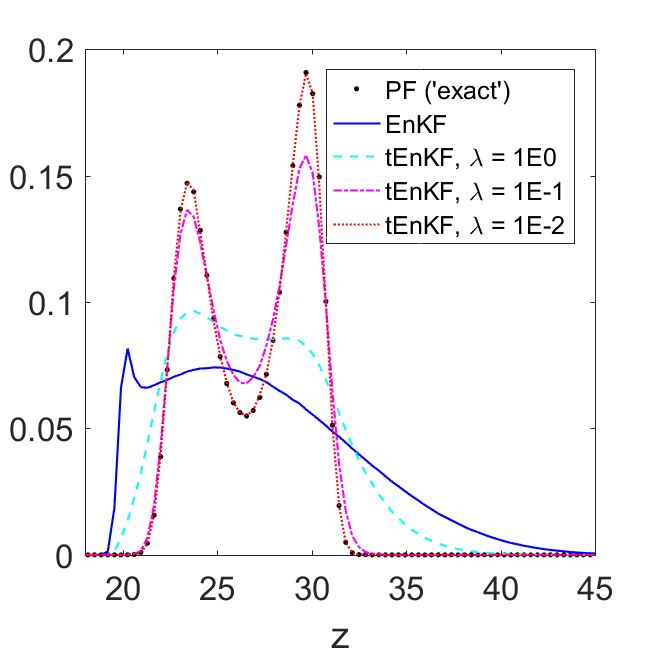}
\begin{tabular}{c}
\vspace{0pt}
\begin{tabular}[t]{|ccc|ccc|c|ccc|cc|cc|}
    $\alpha$ & $\rho$ & $\beta$ & $x_{1,0}$ & $x_{2,0}$ & $x_{3,0}$ & $n$ & $\sigma_{1,0}$ & $\sigma_{2,0}$ & $\sigma_{3,0}$ & $\sigma$ & $\tau$ & $t_1$ & $\Delta t$ \\ \hline
    $10$ & $28$ & $8/3$ & $1.5$ & $y_{truth}$ & $25$ & $10^7$ & $0.1$ & $\tau$ & $0.1$ & $0.01$ & $0.2$ & $1$ & $0.01$
\end{tabular}
\end{tabular}
\caption{TEnKF limiting distributions interpolate between those of the EnKF and the PF. (bottom) Parameters for this experiment. \label{fig:LimitDist}}
\end{figure}

We first consider a simple example that intuitively illustrates how we can correct the bias in the EnKF posterior estimate with the trimming procedure. The Lorenz-63 model \cite{lorenz63}, with an additive stochastic noise term to represent model uncertainty, is given by
\begin{equation} \label{eqn:L63SDE}
    \begin{array}{rclcl}
    \displaystyle \frac{\mathrm{d}x_1}{\mathrm{d}t} & = & \alpha(x_2-x_1) & + & \xi_1(t)    \\[.5pc]
    \displaystyle \frac{\mathrm{d}x_2}{\mathrm{d}t} & = & x_1(\rho - x_3) - x_2 & + & \xi_2(t) \\[.5pc]
    \displaystyle \frac{\mathrm{d}x_3}{\mathrm{d}t} & = & x_1 x_2 - \beta x_3 & + & \xi_3(t) 
    \end{array}
\end{equation}
for $t\geq 0$, 
where $\xi_j(t)$, $j = 1, ..., 3$, are independent Gaussian white noise processes with covariance $\sigma^2\delta_{s,t}$ for all $s\ge 0$, where $\delta$ is the Kronecker $\delta$-function. We obtain solution trajectories of the stochastic differential equation (SDE) numerically using the $O(\Delta t)$ strongly/weakly accurate stochastic Heun method, which employs a trapezoidal discretization of the deterministic part of the integral of Eq.~\eqref{eqn:L63SDE}, and an Euler discretization of the stochastic part. The parameters of this model and the following data assimilation problem are given in Figure \ref{fig:LimitDist}.

For certain parameter values, solutions of the deterministic equations ($\sigma = 0$) are known to be sensitive to initial conditions such that small perturbations due to numerical round-off or other disturbances cause \emph{chaotic} trajectories. We consider the Bayesian filtering problem to estimate the distribution of the model state at time $t = t_1$ from a direct, noisy observation of the second component, $y = x_2 + \epsilon$, where $\epsilon\sim\mathcal{N}(0,\tau^2)$. An independent Gaussian prior distribution is assumed for each state variable initial condition, $x_j(0) \sim \mathcal{N}\left( x_{j,0}, \sigma_{j,0}^2 \right)$, $j = 1, ..., 3$.

We solve this data assimilation problem with the EnKF (Section \ref{sec:enkf}), the TEnKF (Algorithm \ref{alg:tr}) for several fixed values of the tuning parameter $\lambda$, and the bootstrapped PF. The latter is known to provide the exact Bayesian filter solution, and a short description of this algorithm is as follows. The forecast ensemble is determined from a Monte Carlo solution to the original Lorenz-63 SDE at $t = t_1$. Then bootstrapped resampling is applied to the forecast ensemble with weights as defined in Subsection \ref{sec:sample_trim} and the trimming function replaced by the likelihood function
$\mathcal{L}(y^i) = \exp\left[ - \left( y^i-d \right)^2 / \left( 2\tau^2 \right)  \right]$.
For each method, a sufficient ensemble size $n$ is used to accurately resolve the limiting ($n\to\infty$) posterior PDFs (Figure \ref{fig:LimitDist}).

The forecast from the nonlinear Lorenz-63 SDE model can be highly non-Gaussian, despite Gaussian assumptions for the model and initial condition uncertainties. This is the case for the Lorenz-63 SDE example in Figure \ref{fig:LimitDist}, where periodically trajectories can randomly switch orbits around one of two different states. If the observations are not sufficiently frequent and accurate, the forecast becomes bimodal. Since the EnKF algorithm measures only the Gaussian component of the forecast, then the joint distribution between the forecast and observations will be inaccurate, and this can skew the posterior distribution (Figure \ref{fig:LimitDist}). The joint correlations are most inaccurate for forecast members furthest from the observations. However, as these members are trimmed away in the TEnKF algorithm, the joint distribution approaches that of the PF for sufficiently large $n$. To wit, as the trimming parameter $\lambda$ shrinks, the limiting posterior PDF of the TEnKF approaches the exact solution determined from the bootstrapped PF.

\subsection{Restoring convergence with ensemble size}

As discussed earlier, nonlinear models introduce non-Gaussian forecast perturbations which can be difficult to correct in the posterior using a Gaussian filter like the standard EnKF. When observations occur less frequently, the increased forecast interval allows the nonlinearities to have a stronger effect. Under such conditions, the EnKF can fail to decrease estimation errors as the ensemble size is increased. On the other hand, a nonlinear/non-Gaussian filter like the TEnKF can restore convergence by trimming outliers and correcting the limiting posterior distribution.

In the examples which follow, we integrate error in posterior estimates over all ensemble members, rather than only the mean of the posterior estimate. Over $N_{\mathrm{rep}}$ runs of the experiment, we analyze the predictive accuracy of the posterior estimates provided by TEnKF by assimilating a time series of $N_t = t_f/\Delta t_{\mathrm{obs}}$ data points spaced $\Delta t_{\mathrm{obs}}$ seconds apart. The error in the posterior estimate can be computed from the root-mean-square distance of the filter ensemble from the true system state, measured over all state dimensions $j = 1, ..., N$. If $x_{j,k}^i$ and $x_{j,k}^t$ are the $j^\mathrm{th}$ elements of the $i^\mathrm{th}$ ensemble member and truth state vectors, respectively, at time $t_k$, then
\begin{eqnarray}
\mathcal{E}_{m,k} &=& \left[ \frac{1}{n}\sum_{i=1}^{n} \frac{1}{N} \sum_{j=1}^{N} \left( x_{j,k}^i - x_{j,k}^t \right)^2 \right]^{1/2}, \label{eqn:rmses}\\
\mathcal{E}_m &=& \left[ \frac{1}{N_t}\sum_{k=1}^{N_t} \mathcal{E}_{m,k}^2 \right]^{1/2}, \label{eqn:rmsea}
\end{eqnarray}
for $m = 1, ..., N_{\mathrm{rep}}$, are samples from the distribution of time-series \eqref{eqn:rmses} and time-averaged \eqref{eqn:rmsea} posterior root-mean-square errors (RMSE).

\begin{figure}[!t] \centering
\subfloat[][]{
\includegraphics[height=2.6in,clip=true,trim=20 5 40 15]{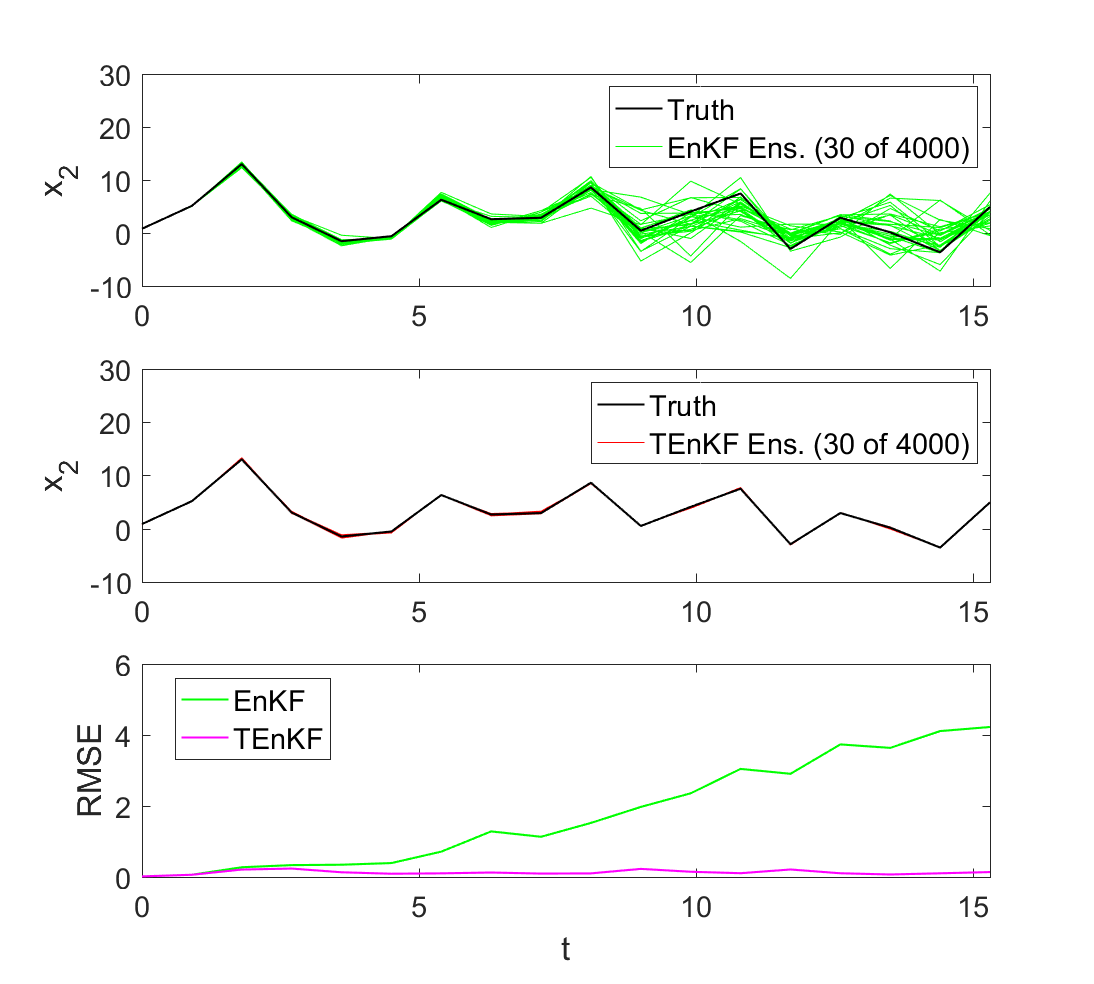}
}
\subfloat[][]{
\begin{tabular}[b]{c|c} \hline
    $N$ & $36$ \\
    $F$ & $8$ \\ \hline
    $t_f$ & $15$ \\
    $\Delta t_{\mathrm{obs}}$ & $0.9$ \\
    $\Delta t$ & $0.01$ \\ \hline
    $\sigma$ & $0.01$ \\
    $\tau$ & $0.05$ \\ \hline
    $n$ & $4000$ \\
    $n_e^\ast$ & $50$ \\ 
    $d_{\mathrm{max}}$ & $3$ \\ \hline
    $\mu_0$ & $1$ \\
    $\mu_1$ & $0.1$ \\
    $\sigma_0$ & $0.01$ \\ \hline
\end{tabular}
}
\caption{(a) EnKF and TEnKF (Algorithm \ref{alg:tr}) performance on noisy Lorenz-96 model over time. (b) Parameters for this experiment. \label{fig:alg1lorenz96traj}}
\end{figure}

To show the applicability of the TEnKF methodology for nonlinear/non-Guassian filters of larger scale, we test Algorithm \ref{alg:tr} on the Lorenz-96 model \cite{lorenz96}. (See Figure \ref{fig:alg1lorenz96traj} for parameters.) This system of nonlinearly-coupled ordinary differential equations is given by
\begin{equation}
\frac{\mathrm{d}x_j}{\mathrm{d} t} = -x_{j-2}x_{j-1} + x_{j-1}x_{j+1} + F  + \xi_j(t)
\end{equation}
for $j = 1, ..., N$ and forcing constant $F>0$. Like the noisy Lorenz-63 model discussed earlier, each noise term $\xi_j(t)$ is an independent Gaussian white noise process with variance $\sigma^2$. (See example in Subsection \ref{subsec:limit_dist_example}.) Lorenz introduced the deterministic version of the Lorenz-96 system ($\sigma = 0$) and showed that for $N = 36$ and $F = 8$, this system exhibits chaotic trajectories. We take direct, noisy observations of the system state at every odd-indexed component, or
\begin{equation}
\mathbf{y} = h(\mathbf{x}) + \mathbf{\epsilon} = \left( x_1, x_3, ..., x_{N-1} \right) + \mathbf{\epsilon}
\end{equation}
with $\mathbf{\epsilon} \sim \mathcal{N}\left( 0 , \tau^2 I_{N/2} \right)$, where $I_s$ is the $s\times s$ identity matrix. In order to prevent degeneracy due to excessive initial errors, we draw the initial condition (IC) from a noisy observation of the truth at time $t = 0$. The truth and unobserved components ($k$ even) of the ensemble member ICs are drawn from the same Gaussian distribution, $\mathcal{N}\!\left( \mu_0 + \mu_1 \cdot z, \sigma_0^2 \right)$, where $z\sim \mathcal{N}\!\left(0,1\right)$ is constant for any given experiment. For the observed components ($k$ odd), the ICs are drawn from the likelihood distribution at $t = 0$ by taking $\mathbf{x}_{2k-1}(0) \sim \mathcal{N}\!\left( \mathbf{y}_{0,k}, \tau^2 \right)$, $k = 1, ..., N/2$.

\begin{figure}[t!] \centering
\includegraphics[height=1.9in,clip=true,trim=5 0 30 15]{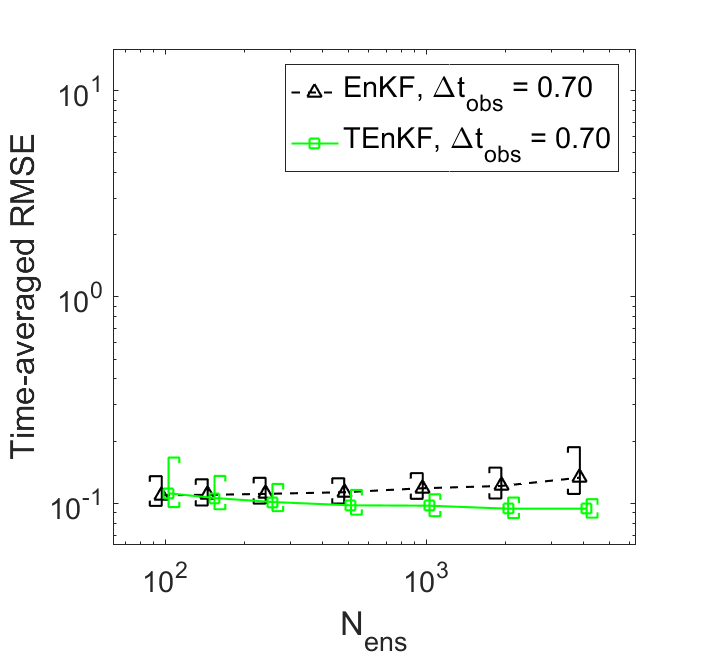}
\includegraphics[height=1.9in,clip=true,trim=5 0 25 15]{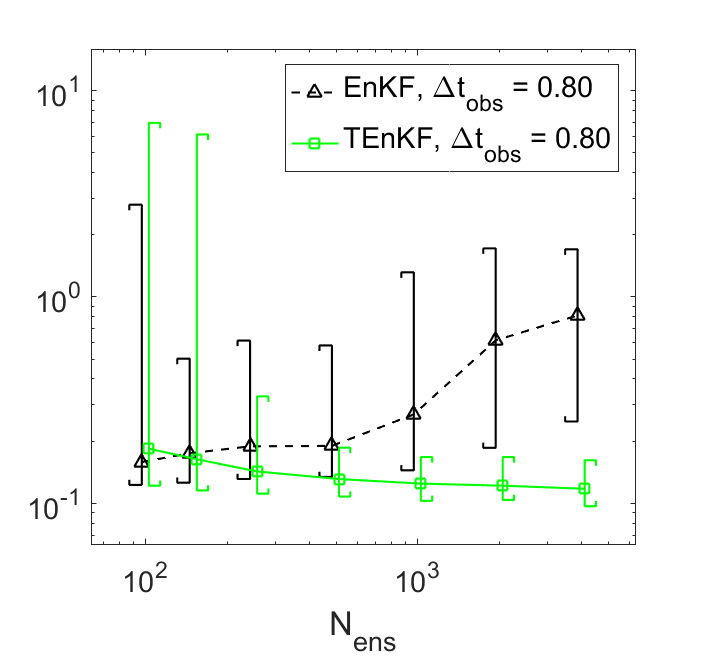}
\includegraphics[height=1.9in,clip=true,trim=5 0 25 15]{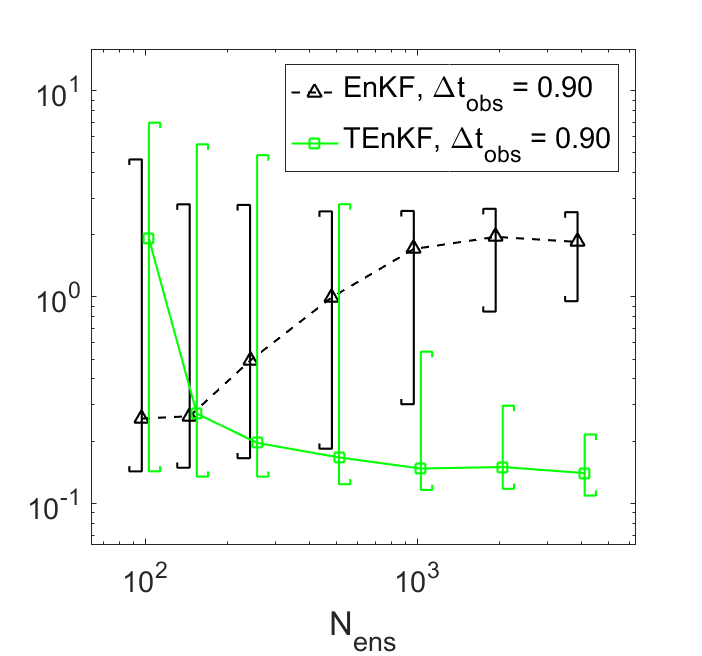}
\caption{Effect of increasing nonlinearity ($\Delta t_{obs}\uparrow$) on the median and inter-quartile range of the time-averaged RMSE errors from the EnKF and TEnKF applied to the noisy Lorenz-96 model over 500 runs. Parameters are given in Figure \ref{fig:alg1lorenz96traj}. \label{fig:alg1lorenz96errors}}
\end{figure}

The TEnKF resolves the non-Gaussian posteriors shaped by the nonlinear model when the standard EnKF cannot (Figure \ref{fig:alg1lorenz96traj}). For the TEnKF, the ensemble size is sufficient to find the proper correlations between observation errors and corrections to the forecast ensemble. In the EnKF, these correlations appear to be spurious and weakened, leading to a false increase in forecast variance that, despite a large ensemble and accurate observations, does not enable the filter to keep track of the truth trajectory. We see these relative differences in tracking proficiency persist over many repetitions of the experiment with different truth realizations (Figure \ref{fig:alg1lorenz96errors}). For observation intervals less than $\Delta t \leq 0.70$, the forecast is sufficiently Gaussian that the performance improvement from the TEnKF is incremental. However, over longer observation intervals the median prediction error is smaller for the TEnKF with any ensemble size larger than $200$. The uncorrected bias from non-Gaussian outliers spurred by the model nonlinearities degrades the EnKF posterior estimate, which worsens as $\Delta t_{\mathrm{obs}}$ is increased. This bias cannot be corrected simply by increasing the ensemble size, and more accuracy is regained by trimming these outliers than is lost due to a smaller effective sample size.

\subsection{Adapting to varying nonlinearity}
\label{subsec:nonlin}

The adaptive procedure in Algorithm \ref{alg:tr} maintains a minimum effective sample size, but it risks over-trimming whenever the non-linearities in the model are weak. The algorithm also risks under-trimming if the forecast distribution requires an ensemble larger than we initially used, since each ensemble member carries more weight and  is less dispensable. The adaptive ensemble size procedure described in Subsection \ref{subsec:AdaptEns} enables the forecast to be enlarged to meet a target effective sample size. We apply this modification to the Lorenz-96 problem. In particular, we consider a deterministic version of the system ($\sigma = 0$), and accelerate forecast generation with a Runge-Kutta 4\textsuperscript{th}-5\textsuperscript{th} adaptive integration scheme. 
The distance measure in \eqref{eq:n_d} is chosen to be
$d(\mathbf{y} , \mathbf{y}^*)=\max_j |y_j - y^*|$.
To compute the new forecast members, we perturb the new initial conditions with $\mathcal{N}\!\left( 0, \sigma_{\mathrm{p}} \right)$ noise independently in each dimension.

\begin{figure}[t!] \centering
\includegraphics[height=2.3in,clip=true,trim=10 0 30 20]{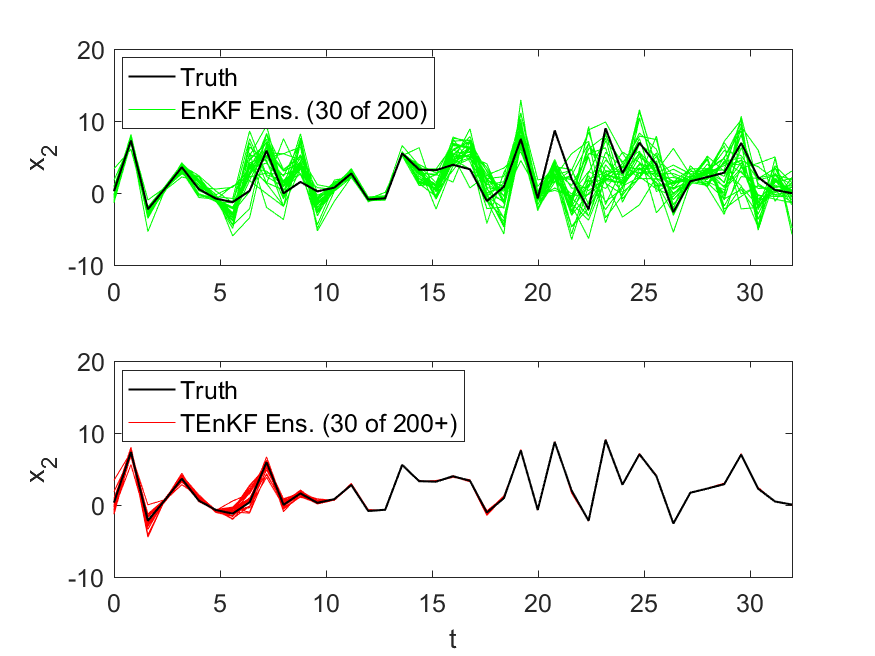}
\includegraphics[height=2.3in,clip=true,trim=10 0 30 20]{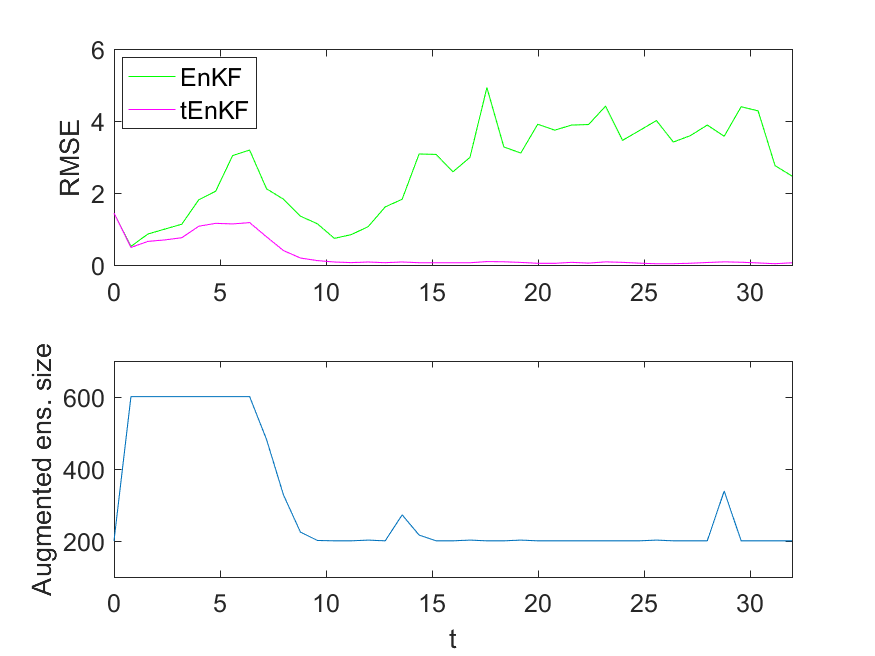} \\
\begin{tabular}[t]{|cc|ccc|ccc|ccc|ccc|}
    $N$ & $F$ & $t_f$ & $\Delta t_{\mathrm{obs}}$ & $\Delta t$ & $n$ & $r_{\mathrm{max}}$ & $d_{\mathrm{max}}$ & $\sigma$ & $\tau$ & $\sigma_p$ & $\mu_0$ & $\mu_1$ & $\sigma_0$ \\ \hline
    $36$ & $8$ & $32$ & $0.8$ & adaptive & $200$ & $3$ & $3$ & $0.01$ & $0.05$ & $0.4$ & $1$ & $0.1$ & $0.01$
\end{tabular}
\caption{EnKF and TEnKF (with adaptive ensemble augmentation) performance on deterministic L96 model. (left) For state component $2$, (bottom) parameters for this experiment. \label{fig:alg2lorenz96traj}}
\end{figure}

Our exhibit of one experimental run shows how the variation of ensemble sizes allow the TEnKF to adjust to time-varying levels of nonlinearity (Figure \ref{fig:alg2lorenz96traj}). Without model error to disperse the forecast ensemble, the standard EnKF is more sensitive to occasional reductions in the effective forecast sample size. Since the TEnKF only augments the ensemble when the effective sample size decreases below $n=200$, we see that typically this ensemble size is sufficient for a Gaussian filter to track the the truth trajectory. However, at several points there are sufficient non-Gaussian outliers generated by the nonlinear model and initial variance to prevent the EnKF from making correct forecast updates, despite having relatively accurate measurements. By allowing the ensemble to enlarge up to $3n$, and then disposing of enough corrupting outliers from the non-Gaussian forecast, the TEnKF is capable of tracking the nonlinear Lorenz 96 model without much more effort over time than the standard EnKF.

\begin{figure}[t!] \centering
\includegraphics[height=3in,clip=true,trim=10 0 20 0]{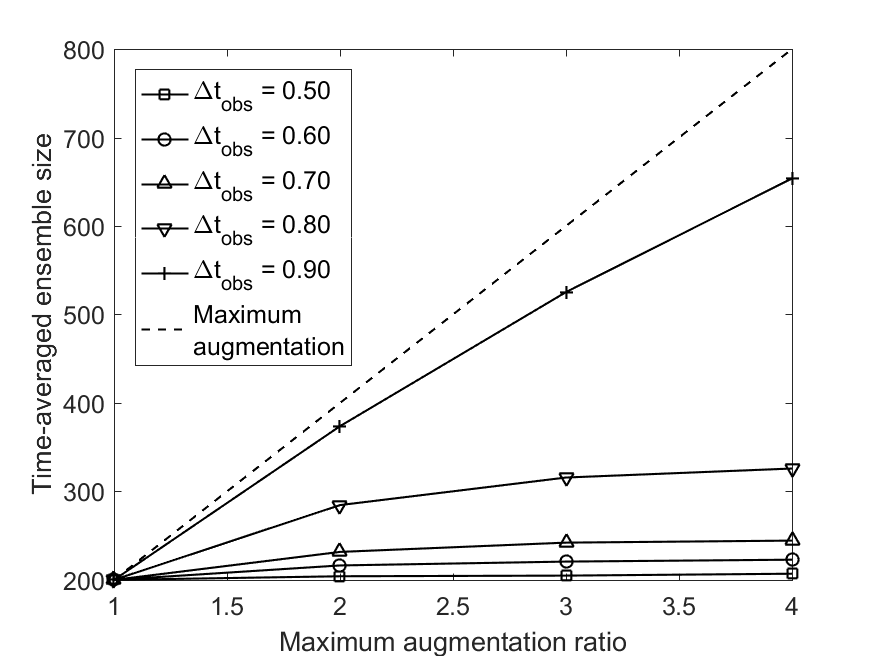}
\caption{Effect of increasing nonlinearity ($\Delta t_{\mathrm{obs}}$) on the time-averaged increase in ensemble size ($n_{aug}/n$) by the TEnKF with adaptive ensemble augmentation, over $250$ repetitions of the experiment. Parameters in Figure \ref{fig:alg2lorenz96traj}. \label{fig:alg2lorenz96effort}}
\end{figure}

To assess the impact of adaptive ensemble augmentation, we examine the rate at which nonlinearities force the TEnKF to augment the ensemble size over several repetitions of the deterministic Lorenz-96 experiment (Figure \ref{fig:alg2lorenz96effort}). 
An ensemble size of $n=200$ within a observation threshold distance of $d_{\mathrm{max}}=3$ from the measurements was sufficient to enable the filter to track the deterministic Lorenz-96 system.
To ensure a robust algorithm, the maximum augmentation ratio, $r_{\mathrm{max}}$, need only be so large that the relative increase in average ensemble size with respect to $r_{\mathrm{max}}$ is negligible. 
At prediction times of $\Delta t_{\mathrm{obs}} = 0.80$, then $r_{\mathrm{max}} \approx 3$. 
Even though at this strength of nonlinearity the typical forecast ensemble size is $n_{\mathrm{aug}}\approx 330$, a capacity of $n r_{\mathrm{max}} \approx 600$ ensemble members is needed to handle the transient nonlinearities while maintaining the effective ensemble size. 
The computational cost increases rapidly with forecast length.
An exponential extrapolation of the $\Delta t_{\mathrm{obs}} = 0.90$ case suggests an average augmented ensemble size as large as $1500$, and a required capacity greater than $4000$ members. 
However, increasing the minimum ensemble size may temper these requirements to some degree.

\section{Conclusions}

We have introduced a trimmed ensemble Kalman filter, or TEnKF, developed as an extension of the ensemble Kalman filter to solve sequential nonlinear non-Gaussian Bayesian inverse problems. This algorithm uses a ``trimming'' function to identify outliers in the observed forecast which contribute to errors in the correlation between the forecast and likelihood ensembles when the forecast is significantly non-Gaussian. For specific trimming functions, we show the TEnKF accurately reproduces the limiting distributions of both the standard EnKF and a particle filter with bootstrapped resampling (i.e., the exact Bayesian posterior) on a non-linear, non-Gaussian test problem. A one-parameter family of trimming functions allows us to interpolate between these limiting distributions to balance adaptively accuracy or efficiency.

An implementation of the TEnKF methodology is presented with adaptive control of the trimming function and the effective ensemble size. Through numerical experiments on stochastic versions of the 3-dimensional Lorenz-63 model and the 36-dimensional Lorenz-96 model, we show the methods restore convergence to the true posterior in cases when the EnKF fails to converge. We also extended the TEnKF to use adaptive ensemble augmentation to overcome transient increases in model nonlinearity and improve efficiency over smoother intervals.

The efficiency results and the flexibility of the TEnKF algorithm present an opportunity to integrate other methods that increase the number of significant ensemble members. For example, consider importance sampling and steering techniques which draw ensemble members toward observations. We may further improve accuracy in uncertainty calculations by combining these with the TEnKF to further increase the effective ensemble size, and then remove any non-Gaussian outliers.

\appendix
\section{Proof of Proposition \ref{proposition:enkf asymptotic general}}
\label{app:proof1}
To derive the PDF of $\tilde{\mathbf{X}}$, we consider the mapping from $(\mathbf{X}, \mathbf{Y})$ to $(\tilde{\mathbf{X}}, \mathbf{Y})$. The Jacobian matrix of this mapping is
\[ 
J=
\left[ \begin{array}{cc}
I & -\mathcal{K}\\
0 & I \end{array} \right],
\]
where $I$ is the identity matrix. It is easy to verify that the determinant of $J$ is $|J| = 1$. So, we have the joint PDF of $(\tilde{\mathbf{X}}, \mathbf{Y})$:
\begin{equation}
\label{eq:pdf_transform}
p_{\tilde{\mathbf{X}} \mathbf{Y}} ( \tilde{\mathbf{x}}, \mathbf{y})
= p_{\mathbf{X} \mathbf{Y}} ( \mathbf{x}, \mathbf{y}) |J|^{-1} 
= p_{\mathbf{X} \mathbf{Y}} ( \mathbf{x}, \mathbf{y}).
\end{equation}
Using the product rule $p_{\mathbf{X} \mathbf{Y}} ( \mathbf{x}, \mathbf{y}) = p_{\mathbf{X} | \mathbf{y}} ( \mathbf{x}) p_{\mathbf{Y}} (\mathbf{y})$, and substituting the relationship $\mathbf{x} = \tilde{\mathbf{x}} - \mathcal{K} (\mathbf{y}^* - \mathbf{x})$ to Eq.~\eqref{eq:pdf_transform}, we have
\begin{equation}
p_{\tilde{\mathbf{X}} \mathbf{Y}} ( \tilde{\mathbf{x}}, \mathbf{y})
= p_{\mathbf{X} | \mathbf{y}} ( \tilde{\mathbf{x}} - \mathcal{K} (\mathbf{y}^* - \mathbf{x}) ) p_{\mathbf{Y}} (\mathbf{y}).
\end{equation}
Finally, by integrating out $\mathbf{y}$, we have the marginal PDF of $\tilde{\mathbf{X}}$:
\[
p_{\tilde{\mathbf{X}}} (\tilde{\mathbf{x}})= 
\int 
p_{\mathbf{X}|\mathbf{y}} (\tilde{\mathbf{x}} - \mathcal{K} (\mathbf{y}^* - \mathbf{y}))
p_{\mathbf{Y}}(\mathbf{y})
d\mathbf{y}.
\]
\hfill $\Box$

\section{Proof of Proposition \ref{proposition:tenkf asymptotic general}}
\label{app:proof2}
By Proposition \ref{proposition:enkf asymptotic general}, we have
\begin{equation}
\label{eq:proof1}
p^t_{\tilde{\mathbf{X}}} (\tilde{\mathbf{x}})=
\int 
p^t_{\mathbf{X}|\mathbf{y}} (\tilde{\mathbf{x}} - \mathcal{K} (\mathbf{y}^* - \mathbf{y}))
p^t_{\mathbf{Y}}(\mathbf{y})
d\mathbf{y},
\end{equation}
where $p^t_{\mathbf{X}|\mathbf{y}}(\cdot)$ and $p^t_{\mathbf{Y}}(\cdot)$ are the conditional PDF and the marginal PDF determined by the adjusted joint PDF \eqref{eq:adjust_joint}.

By the definition of \eqref{eq:adjust_joint} and the sum rule (i.e., the marginal PDF can be calculated by summing/integrating the joint PDF over other random variables), we have
\begin{equation}
\label{eq:proof2}
p^t_{\mathbf{Y}}(\mathbf{y})
=
\int p^t_{\mathbf{XY}}(\mathbf{x}, \mathbf{y}) d\mathbf{x}
=
\int c_t t(\mathbf{y}) p_{\mathbf{XY}}(\mathbf{x}, \mathbf{y}) d\mathbf{x}
=
c_t t(\mathbf{y}) \int  p_{\mathbf{XY}}(\mathbf{x}, \mathbf{y}) d\mathbf{x}
=
c_t t(\mathbf{y}) p_{\mathbf{Y}}(\mathbf{y}).
\end{equation}

By Eq.~\eqref{eq:adjust_joint}, Eq.~\eqref{eq:proof2}, and the fact that the conditional PDF is equal to the joint PDF divided by the marginal PDF, we have
\begin{equation}
\label{eq:proof3}
p^t_{\mathbf{X|y}} (\mathbf{x}) 
=
\frac{p^t_{\mathbf{XY}}(\mathbf{x}, \mathbf{y})}{p^t_{\mathbf{Y}}(\mathbf{y})}
=
\frac{c_t t(\mathbf{y}) p_{\mathbf{XY}}(\mathbf{x},\mathbf{y})}{c_t t(\mathbf{y}) p_{\mathbf{Y}}(\mathbf{y})}
=
\frac{p_{\mathbf{XY}}(\mathbf{x},\mathbf{y})}{p_{\mathbf{Y}}(\mathbf{y})}
=
p_{\mathbf{X|y}} (\mathbf{x}). 
\end{equation}

Finally, combining Eqs.~\eqref{eq:proof1}, \eqref{eq:proof2} and \eqref{eq:proof3} yields
\[
p^t_{\tilde{\mathbf{X}}} (\tilde{\mathbf{x}})=
\int 
p_{\mathbf{X}|\mathbf{y}} (\tilde{\mathbf{x}} - \mathcal{K} (\mathbf{y}^* - \mathbf{y}))
p_{\mathbf{Y}}(\mathbf{y}) c_t t(\mathbf{y})
d\mathbf{y}.
\]
\hfill $\Box$

\bibliographystyle{siamplain}
\bibliography{references}
\end{document}